%TeX MACRO for American Institute of Physics conference proceedings

\magnification=\magstep1
\hoffset=0.5truein
\hsize=5.75truein
\vsize=8.75truein
\baselineskip=12.045truept   % sets baseline spacing to 6 lines per inch %
\parindent=2.5em	     % sets paragraph indent to 5 spaces %
\parskip=0pt
\def\hi{\noindent \hangindent=2.5em}
\def\bigskip{\vskip 12.045pt}		% skip one line %
\overfullrule=0pt
\font\twelverm=cmr10 at 12truept 	% sets type size to 12 pt %
\twelverm				% document font 12 point Roman %
\nopagenumbers                          % page numbers inserted later
% 
% Definitions
% 
\def\title#1{\centerline{\bf #1}}
\def\author#1{\bigskip\centerline{#1}}
\def\address#1{\centerline{#1}}
\def\sec#1{\bigskip\centerline{#1}\bigskip}

%%
% some useful but optional definitions 
%%

\def\ie{{\it i.e.\ }}

\def\go{
\mathrel{\raise.3ex\hbox{$>$}\mkern-14mu\lower0.6ex\hbox{$\sim$}}
}
\def\lo{
\mathrel{\raise.3ex\hbox{$<$}\mkern-14mu\lower0.6ex\hbox{$\sim$}}
}
\def\Oigm{\Omega_{\rm igm}}
\def\Ho{H_0}
\def\J24{J_{24}}
\def\T4{T_4}
\def\Obbn{\Omega_{\rm bbn}}

%%%%%%%%%%%%%%%%%%%%%%%%%%%%%%%%%%%%%%%%%%%%%%%%%%%%%%%%%%%%%%%%%%%%%%%%%%%%

\bigskip
\title{THE PHYSICAL STATE OF THE INTERGALACTIC MEDIUM}
\title{OR}
\title{CAN WE MEASURE $Y$?}
\author{James G. Bartlett, Domingos Barbosa, Alain Blanchard}
\address{Observatoire de Strasbourg, U.L.P., 11, Rue de l'Universit\'e,
	  67000 Strasbourg, FRANCE}
\address{Unit\'e associ\'ee au CNRS} 
\address{bartlett@astro.u-strasbg.fr}

\sec{ABSTRACT}

	We present an argument for a {\it lower limit} to the 
Compton-$y$ parameter describing spectral distortions of the
cosmic microwave background (CMB).  The absence of a detectable 
Gunn-Peterson signal in the spectra of high redshift quasars
demands a high ionization state of the intergalactic medium (IGM).
Given an ionizing flux at the lower end of the 
range indicated by the proximity effect, 
an IGM representing a significant fraction
of the nucleosynthesis-predicted baryon density must 
be heated by sources other than the photon flux to 
a temperature $\go {\rm few} \times 10^5\, K$.  
Such a gas at the redshift of the highest observed
quasars, $z\sim 5$, will produce a $y\go 10^{-6}$.  
This lower limit on $y$ rises if the Universe is 
open, if there is a cosmological constant, or if one adopts 
an IGM with a density larger than the prediction of standard 
Big Bang nucleosynthesis.

\sec{1. Introduction}

	Arriving from high redshift, the photons of the cosmic 
microwave background (CMB) carry information on the state of the
material through which they have traveled, in particular the 
intergalactic medium (IGM).  Any interaction at low redshifts
between the IGM and the photons leads to a characteristic 
spectral distortion of the CMB commonly referred to as the 
{\it Compton $y$-distortion} (Zel'dovich \& Sunyaev 1969), 
due to its physical origin 
in Compton scattering between electrons and photons and the 
fact that it is described by a single number called
the $y$ parameter (defined below).  The deviation from 
a pure blackbody appears in the Wien tail of the CMB, in the far 
infrared at wavelengths from about 1 mm to 100 microns.  
Observations of this background thus provide constraints on 
the density and temperature of the ionized component of the IGM; 
an upper limit on $y$ places {\it upper} limits on the temperature 
and density.  The current limit from FIRAS is $y<2.5\times 10^{-5}$ 
(Mather et al. 1994).  

	Studies of quasar spectra also provide information on the 
IGM: the absence of a detectable Ly$\alpha$ absorption trough 
indicates that the IGM contains extremely little neutral Hydrogen 
(Gunn \& Peterson 1965).  This so-called Gunn--Peterson (GP) 
test places stringent limits on the quantity of HI found in the 
IGM since a redshift of about 5 and forces us to conclude 
that either their is no uniformly distributed 
IGM or that it is very highly ionized.  For an assumed total 
(non-zero) IGM density, the GP test thus provides a 
{\it lower} limit to the temperature of the ionized component.   

	As these two types of observations constrain the IGM
in the temperature-density plane in {\it opposite} directions, one
may hope to draw interesting conclusions concerning the physical
state of the IGM by combining the two types of studies.  This 
is what we shall explore in this contribution by asking the 
question ``given the GP limits, what is the corresponding 
lower bound to $y$?''

\sec{2. The Observables}

	The two observational quantities under consideration 
are the optical depth to Ly$\alpha$ absorption and the Compton  
parameter $y$.  For an observed frequency $\omega_0$,
the former may be written as an integral along the
photon path, $\omega(z) = \omega_0 (1+z)$, 
up to the redshift $z_{\rm qso}$ of the quasar 
under observation (we write our equations in units for 
which $c=k=h/2\pi=1$):
$$ 
\tau [\Oigm,T,\J24;\Ho,\Omega,\Lambda] = \int_0^{z_{\rm qso}}
	dz |{{dt}\over{dz}}| n_{\rm HI}(z) \sigma_\alpha[\omega(z)], 
			\eqno(1a) 
$$
where $\sigma_\alpha(\omega)$ is the absorption cross-section;
$n_{\rm HI}$ is the density of neutral Hydrogen; $\Oigm$ is the 
mass density of the IGM relative to the critical
density -- $n_{\rm igm} = \Oigm (3\Ho^2/8\pi G)/m_{\rm p}$, 
$m_{\rm p}$ being the proton mass (in our simple calculation
we consider only Hydrogen in the IGM); $T$ is the temperature of the 
IGM and $\J24$ is the photon flux at the Hydrogen 
ionization threshold in units of $10^{-24}\;$ W/m$^2$ Hz ster 
(in more sane units of cgs, and not the PC units enforced 
during the conference, this corresponds to the 
familiar $J_{21}$!).  Studies of the proximity effect around quasars
suggest that $\J24 \sim 0.3-3$ (Bajtlik et al. 1988), although
estimates of the quasar contribution to the ionizing background
favor values at the low end of (in fact below) this range
(Madau 1992).  The function 
$|{{dt}\over{dz}}|$ contains all of the cosmological parameters: $\Omega$,
the total mass density of the universe; $\Ho$, the Hubble constant 
(hereafter $\Ho = 100h\; $km/s/Mpc); and $\Lambda$, a possible 
cosmological constant.  If we restrict ourselves for a 
moment to the case of a critical universe ($\Omega = 1$), 
this expression reduces to 
$$  
\tau = 4.4\times 10^5 (1-\chi) (\Oigm h^2) h^{-1} (1+z)^{1.5}, \eqno(1b) 
$$
in which appears the {\it ionized} H fraction $\chi(\Oigm,T,\J24)$.
This ionized fraction is defined as $n_{\rm p}/n_{\rm igm}$, where
$n_{\rm p}$ is the density of free protons.  
We have explicitly written the dependence on IGM density in terms of
$\Oigm h^2$ for comparison with the predictions of Big Bang 
nucleosynthesis theory.  This theory constrains the number 
$\Obbn h^2 \sim 0.013$ (Walker et al. 1991).
We see immediately that for any non-negligible quantity of intergalactic
gas, $\chi$ must be very close to unity to avoid producing an observable 
amount of absorption, \ie the medium must be highly ionized.
This in turn implies, as we will quantify in the following
section, a high temperature, particularly for the low fluxes 
$\J24$ suggested by integration of quasar emission.  For our
purposes, the most useful limits on $\tau$ have been given by
Giallongo et al. (1994): $\tau < 0.05$ at $z=4.3$.

	The Compton parameter is defined as the integral along 
the line-of-sight of the Compton optical depth times an effective 
energy transfer coefficient $T/m_{\rm e}$, where $m_{\rm e}$ is the
electron mass:
$$ 
y[\Oigm,T;\Omega,\Ho,\Lambda] = \int_{0}^{z_{\rm qso}} dz |{{dt}\over{dz}}|
	\bigg( {T\over{m_{\rm e}}} \bigg) \chi n_{\rm igm} \sigma_{\rm T}, 
		\eqno(2a)
$$
which for a critical universe becomes ($\sigma_{\rm T}$ being the Thompson
cross-section)
$$  y = 10^{-6} \T4 (\Oigm h^2) h^{-1} \bigg({{1+z}\over {6}} \bigg)^{1.5}.
		\eqno(2b)
$$
In the second expression we have defined $\T4$ as $T/10^4$ K 
and assumed that $\chi=1$.  Notice also that the numerical 
value is referred to a redshift of 5.  

	From these formulae, we easily see the physics helping us to
constrain the IGM.  A limit on $\tau$ eliminates high density, 
{\it low} temperature regions of the IGM temperature-density plane while
limits on $y$ eliminate high density, {\it high} temperature regions of this
phase space (figures 1 and 2).  Thus we ``squeeze'' the IGM into 
the left-most portions of the diagram, towards low total IGM density.
The redshift dependence in equations (1) and (2) provides motivation
to search for the GP effect at the highest possible redshifts, 
although one must remember that the increasing Ly$\alpha$ forest
line density makes this a more and more difficult proposition 
(beyond the primary difficulty of finding such high redshift 
quasars).  All of this assumes that we know the dependence of $\chi$ 
on the density and temperature of the IGM, or, in other words, that we 
know the ionization mechanism.  In the following section we consider 
ionization by collisions and radiation.   

	A remark on the dependence of the constraints on the 
cosmological model, which enters only through the expansion time
$dt/dz$ -- All models for which $\Omega<1$, with or without a 
cosmological constant, result in tighter constraints on the IGM:
The slower deceleration in these scenarios implies a longer expansion
time at any given redshift $z$ ($dt/dz$ at any given 
$z$ increases relative to the critical case) and therefore 
larger optical depths to {\it both} Ly$\alpha$ absorption 
and Compton scattering.  Lowering the Hubble constant for
a fixed, physical IGM density, $\Oigm h^2$, has the 
same effect. In both case the reason is the same -- the Universe 
is older.

\sec{3. Constraints}

	To proceed and actually constrain the IGM, 
we must model the ionization physics of the gas, in order to
find the functional form of $\chi(\Oigm,T,\J24)$.
Assuming ionization equilibrium maintained by
a flux of ionizing photons and by collisions, we write
$$
	\alpha n_{\rm p}^2 = \Gamma_{\rm pi} n_{\rm HI}
		+ \Gamma_{\rm c} n_{\rm p} n_{\rm HI}, \eqno(3)
$$
where $\alpha$ is the recombination rate; $\Gamma_{\rm pi}$, the
photoionization rate, dependent on $\J24$; and 
$\Gamma_{\rm c}$, the collisional ionization rate.  
Recall that we consider only Hydrogen in our simple calculation.
The assumption of ionization equilibrium eliminates any dependence
on the state of the medium prior to the redshift under 
consideration.  Although
the recombination rate is smaller than the expansion rate at
redshifts less than about 5, this assumption seems reasonable because
even the small quantity of gas capable of recombining in 
one expansion time would quickly violate the GP limits if there
were no compensating source of ionization.  

	To get a feel for the numbers, consider the case of 
pure photoionization in a critical Universe with $h=1/2$.  
Putting $\Gamma_{\rm c} = 0$, $\Gamma_{\rm pi} \sim 4\times 10^{-12} 
\J24\; $s$^{-1}$ and $\alpha \sim 4\times 10^{-13} \T4^{-0.7}\; $
cm$^3\; $s$^{-1}$ (Peebles 1993) in equation (3), one finds 
$(1-\chi)/\chi^2$, and, with the approximation that $\chi\approx 1$,
one then arrives (Eq. 1b) at $ \tau = 2\times 10^3 (\Oigm h^2)^2 h^{-1} 
\T4^{-0.7} \J24^{-1} [(1+z)/6]^{4.5}$.  
Thus, in order to satisfy $\tau < 0.05$ at a redshift of 5, 
an IGM with a density $\sim \Obbn = 0.013/h^2$ subjected to an 
ionizing flux $\J24\sim 1$ must have a temperature of $\T4\go 50$.
This same gas would then produce a $y\go 10^{-6}$.  
Results of a more general and careful
calculation, taking into account both photoionization and 
ionization by collisions, are shown in the figures.

\sec{4. Conclusion}

	The goal of our discussion is to show the nature of
the constraints imposed on the IGM by combining the GP test with 
limits on the $y$ parameter.  The quantitative results are given 
in the figures for the current GP and $y$ limits, as well as for an
eventual limit of $y<10^{-6}$.  Spectral observations of the 
CMB with the latter sensitivity would either provide the evidence
for or eliminate the possibility of hiding a large quantity
of baryons in the IGM.  This would be an important result 
for some cosmological scenarios, such as the PIB model (Peebles 1987a;
1987b), and of special interest in light of the cluster ``baryon crisis'' 
(White et al. 1993).  At a level of $y\sim 10^{-6}$, one is at the
threshold of eliminating an IGM containing most of the 
baryons predicted by standard nucleosynthesis, particularly 
for low density Universes (figure 2).  Such a conclusion
would have important implications for the nature of the dark
matter in galactic halos.  There is the additional hope of
improving the GP limits with high resolution 
spectra of quasars at ever larger redshifts.  A combined effort 
of this kind would permit us to draw interesting and important 
conclusions on the history of the baryonic content of the 
Universe during galaxy formation.

%\sec{Acknowledgements}

%This section thanks all those people who have not been thanked,
%whatever the reason.

\sec{References}

\hi{Bajtlik, S., Duncan, R.C., \& Ostriker, J.P. 1988, ApJ, 327, 570}

\hi{Giallongo, E., D'Odorico, S., Fontana, A. et al. 1994, ApJ, 425, L1}

\hi{Gunn, J.E., \& Peterson, B.A. 1965, ApJ, 142, 1633}

\hi{Madau, P. 1992, ApJ, 389, L1}

\hi{Mather, J.C. et al. 1994, ApJ, 420, 439}

\hi{Peebles, P.J.E. 1993, Principles of Physical Cosmology, 
	Princeton University Press (Princeton:New Jersey)}

\hi{Peebles, P.J.E. 1987a, ApJ, 315, L73}

\hi{Peebles, P.J.E 1987b, Nature, 327, 210}

\hi{Walker, T.P., Steigman, G., Kang, H-S, Schramm, D.M., \&
	Olive, K.A. 1991, ApJ, 376, 51}

\hi{White, S.D.M., Navarro, J.F., Evrard, A.E., Frenk, C.S. 1993,
	Nature, 366, 429}

\hi{Zel'dovich, Ya. B., \& Sunyaev, R.A. 1969, Ap\&SS, 4, 301} 

%\hi{NoName, W. 1993, ApJ, 666, 666}

%\hi{NoName, Jr., W. 1993, ApJ, 666, 667}

%%
% Printing instructions for VMS-based systems:
% 
%   $tex filename.tex   % creates filename.dvi (DeVice-Independent file)
%
%   $dviqms8 filename   % creates PostScript output
%
%      This makes filename.bit (where $dviqms8 :== $tex$:dvilg8), 
%      an output file which can be printed on a QMS laser printer.
%
%   $print/que=your_qms_printer filename.bit
%
% It may be necessary to initialize the laser printer first.
%
% The proper command for the last step depends on the type of printer,
% so ask your colleagues or system manager.
%%
%%
% Printing instructions for Unix systems:
% 
%   >tex filename.tex   % creates filename.dvi (DeVice-Independent file)
%
%   >dvips filename     % creates PostScript output
%
%      This makes filename.ps    an output file which can be printed 
%      on a laser printer.
%
%   >lpr filename.ps
%
% It may be necessary to initialize the laser printer first.
%
% The proper command for the last step depends on the type of printer,
% so ask your colleagues or system manager.
%%
vskip 3 truecm

\sec{Figure Captions}

{\bf Figure 1}  The constraints in the temperature-mass plane of the
IGM for a flat universe with a Hubble constant of 50 km/s/Mpc.
The Giallongo et al. (1994) GP limit ($\tau < 0.05$ at $z=4.3$)
eliminates the space below and to the right of the dashed curve.
The calculation accounts for ionization by both collisions and an 
ionizing flux of $\J24=1$.  The change in slope along this
curve at the higher temperatures signals the increasing importance
of collisional ionization.  The dotted-dashed curve shows the 
corresponding boundary for a flux $\J24=0.3$; this flux is
more in line with estimates of the quasar contribution to 
the ionizing background (Madau 1992).  The constraints 
imposed by observations of the CMB are drawn for the actual
FIRAS limit and for a limit of $y<10^{-6}$.  These constraints
eliminate the space above and to the right of the respective 
curves.  Big bang nucleosynthesis bounds the baryon density
to the range indicated by the vertical dotted lines.  
The straight line labeled {\it Pure photoionisation line}
gives the minimum temperature possible for the medium 
at a given $\tau$ (here $=0.05$), corresponding to the case
where the gas is only heated by the ionizing photons; it is
parameterized by $\J24$.  For this reason,
we see the dashed and dotted-dashed curves terminate
at two different points along this line.

\vskip 0.5 truecm

{\bf Figure 2}  Same as figure 1 for $\Omega=0.1$.  Note the
change of scale along the abscissa.

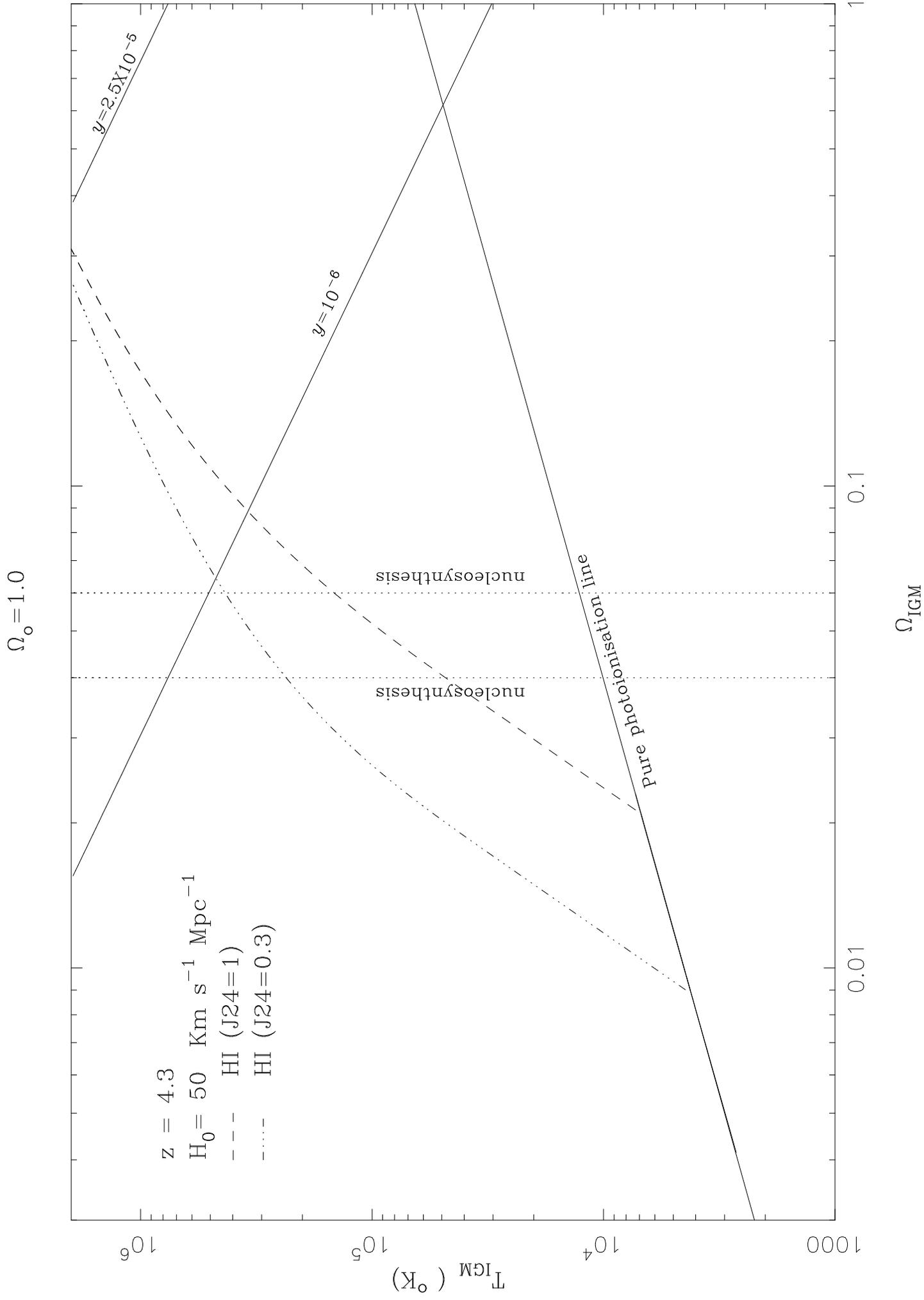
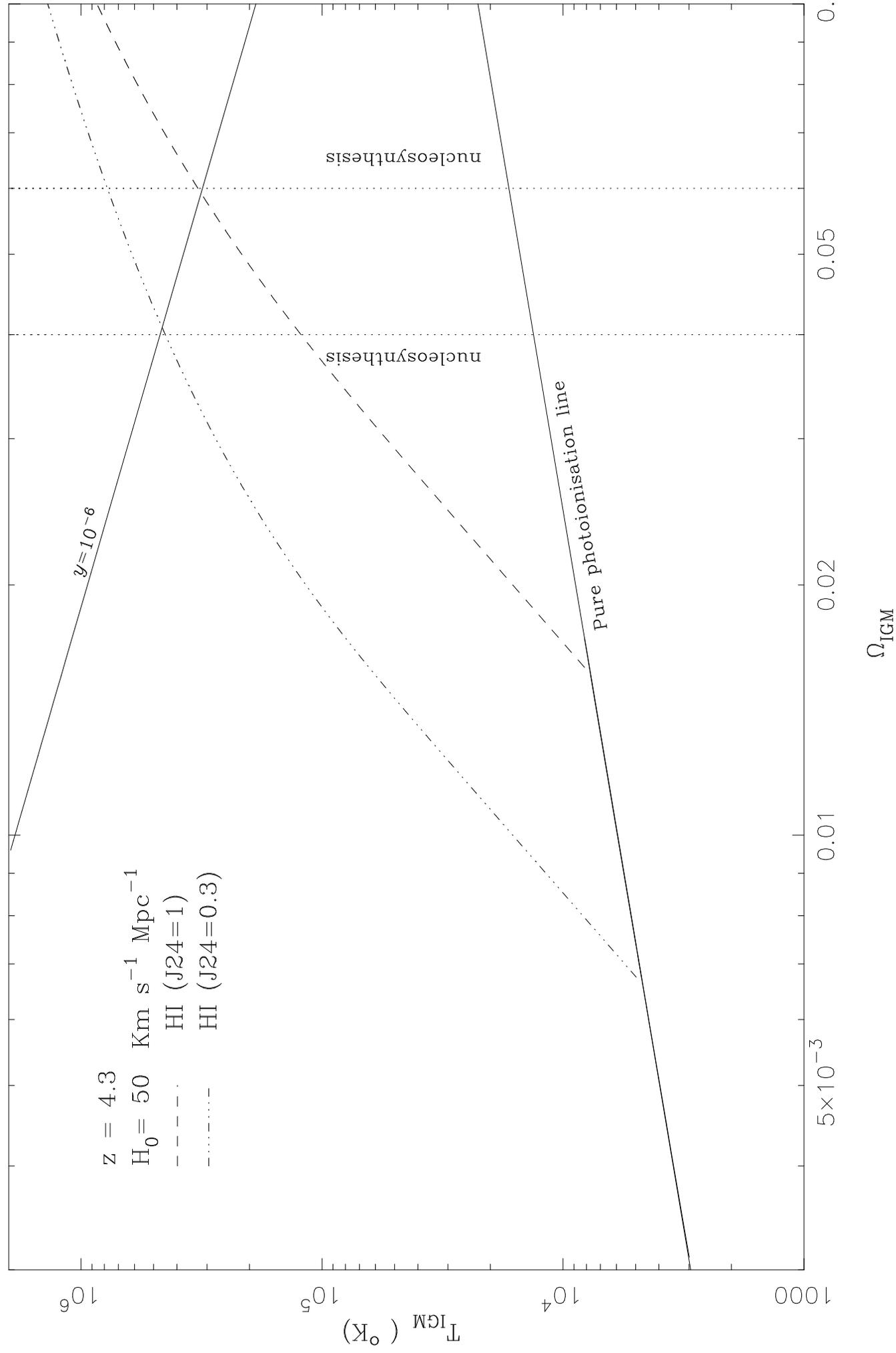
\bye